\documentclass[fleqn,usenatbib]{mnras}

\usepackage{newtxtext,newtxmath}
\usepackage[normalem]{ulem}

\usepackage[T1]{fontenc}

\DeclareRobustCommand{\VAN}[3]{#2}
\let\VANthebibliography\thebibliography
\def\thebibliography{\DeclareRobustCommand{\VAN}[3]{##3}\VANthebibliography}

\def\beq{\begin{equation}}
\def\eeq{\end{equation}}

\usepackage{graphicx}	
\usepackage{amsmath}	

\title[Vector clouds around SgrA$^*$]{Using the motion of S2 to constrain vector clouds around SgrA$^*$}

\author[GRAVITY Collaboration]{GRAVITY Collaboration \thanks{GRAVITY is developed in collaboration by 
MPE, 
LESIA of Paris Observatory / CNRS / Sorbonne Universit\'e / Univ. Paris Diderot 
and IPAG of Universit\'e Grenoble Alpes / CNRS, 
MPIA, 
Univ. of  Cologne,
CENTRA - Centro de Astrof\'{\i}sica e Gravita\c c\~ao, and ESO. Corresponding authors: A.~Foschi 
(arianna.foschi@tecnico.ulisboa.pt) \& P.J.V.~Garcia (pgarcia@fe.up.pt)
}:
A.~Foschi$^{1, 2}$,
R.~Abuter$^{3}$,
K. Abd El Dayem$^{4}$,
N.~Aimar$^{4}$, \newauthor
P.~Amaro Seoane$^{6, 5, 7, 21}$, 
A.~Amorim$^{1, 8}$, 
J.P.~Berger$^{9}$,
H.~Bonnet$^{3}$, 
G.~Bourdarot$^{5}$, 
W.~Brandner$^{10}$, \newauthor 
R.~Davies$^{5}$, 
P.T.~de~Zeeuw$^{11}$, 
D.~Defrère$^{19}$,
J.~Dexter$^{12}$, 
A.~Drescher$^{5}$, 
A.~Eckart$^{16, 18}$,
F.~Eisenhauer$^{5}$, \newauthor
N.M.~F\"orster~Schreiber$^{5}$,
P.J.V.~Garcia$^{1, 2}$, 
R.~Genzel$^{5, 13}$, 
S.~Gillessen$^{5}$, 
T.~Gomes$^{1, 2}$,
X.~Haubois$^{14}$,\newauthor
G.~Hei{\ss}el$^{4, 15}$, 
Th.~Henning$^{10}$,
L.~Jochum$^{14}$,
L.~Jocou$^{10}$, 
A.~Kaufer$^{14}$,   
L.~Kreidberg$^{10}$, 
S.~Lacour$^{4}$, \newauthor
V.~Lapeyr\`ere$^{4}$,  
J.-B.~Le~Bouquin$^{9}$, 
P.~L\'ena$^{4}$, 
D.~Lutz$^{5}$, 
F.~Mang$^{5}$,
F.~Millour$^{20}$,
T.~Ott$^{5}$,
T.~Paumard$^{4}$, \newauthor
K.~Perraut$^{9}$, 
G.~Perrin$^{4}$, 
O.~Pfuhl$^{3, 5}$, 
S.~Rabien$^{5}$, 
D.C.~Ribeiro$^{5}$, 
M.~Sadun Bordoni$^{5}$, 
S.~Scheithauer$^{10}$,  \newauthor
J.~Shangguan$^{5}$,
T.~Shimizu$^{5}$,  
J.~Stadler$^{5, 17}$, 
C.~Straubmeier$^{16}$, 
E.~Sturm$^{5}$, 
M.~Subroweit$^{16}$,
L.J.~Tacconi$^{5}$,  \newauthor
F.~Vincent$^{4}$, 
S.~von~Fellenberg$^{5, 18}$
and J.~Woillez$^{3}$ 
\\
$^{1}$CENTRA - Centro de Astrof\'{\i}sica e
Gravita\c c\~ao, IST, Universidade de Lisboa, 1049-001 Lisboa,
Portugal\\
$^2$Faculdade de Engenharia, Universidade do Porto, rua Dr. Roberto
Frias, 4200-465 Porto, Portugal\\ 
$^3$European Southern Observatory, Karl-Schwarzschild-Stra{\ss}e 2, 85748
Garching, Germany\\
$^4$LESIA, Observatoire de Paris, Universit\'e PSL, CNRS, Sorbonne Universit\'e, Universit\'e de Paris, 5 place Jules Janssen, 92195 Meudon, France\\
$^5$Max Planck Institute for extraterrestrial Physics,
Giessenbachstra{\ss}e~1, 85748 Garching, Germany\\
$^{6}$Universitat Politècnica de València, València, Spain \\
$^{7}$ Kavli Institute for Astronomy and Astrophysics, Beijing, China \\
$^{8}$Universidade de Lisboa - Faculdade de Ci\^encias, Campo Grande,
1749-016 Lisboa, Portugal\\
$^{9}$Univ. Grenoble Alpes, CNRS, IPAG, 38000 Grenoble, France\\
$^{10}$Max Planck Institute for Astronomy, K\"onigstuhl 17, 
69117 Heidelberg, Germany\\
$^{11}$Leiden University, 2311EZ Leiden, The Netherlands\\
$^{12}$Department of Astrophysical \& Planetary Sciences, JILA, Duane Physics Bldg., 2000 Colorado Ave, University of Colorado, Boulder, CO 80309, USA\\
$^{13}$Departments of Physics and Astronomy, Le Conte Hall, University
of California, Berkeley, CA 94720, USA\\
$^{14}$European Southern Observatory, Casilla 19001, Santiago 19, Chile\\
$^{15}$Advanced Concepts Team, European Space Agency, TEC-SF, ESTEC, Keplerlaan 1, 2201, AZ Noordwijk, The Netherlands \\
$^{16}$ $1^{\rm st}$ Institute of Physics, University of Cologne,
Z\"ulpicher Stra{\ss}e 77, 50937 Cologne, Germany\\
$^{17}$Max Planck Institute for Astrophysics, Karl-Schwarzschild-Stra{\ss}e 1, D-85748
Garching, Germany\\
$^{18}$Max Planck Institute for Radio Astronomy, auf dem H\"ugel 69, D-53121 Bonn, Germany \\
$^{19}$Institute of Astronomy, KU Leuven, Celestijnenlaan 200D, 3001 Leuven, Belgium \\
$^{20}$Université C\^{o}te d'Azur, Observatoire de la  C\^{o}te d'Azur, CNRS, Lagrange, France \\
$^{21}$ Higgs Centre for Theoretical Physics, Edinburgh, UK}
\date{Accepted XXX. Received YYY; in original form ZZZ}

\pubyear{2023}

\begin{document}
\label{firstpage}
\pagerange{\pageref{firstpage}--\pageref{lastpage}}
\maketitle

\begin{abstract}
The dark compact object at the centre of the Milky Way is well established to be a supermassive black hole with mass $M_{\bullet} \sim 4.3 \cdot 10^6 \, M_{\odot}$, but the nature of its environment is still under debate. In this work, we used astrometric and spectroscopic measurements of the motion of the star S2, one of the closest stars to the massive black hole, to determine an upper limit on an extended mass composed of a massive vector field around Sagittarius A*. For a vector with effective mass $10^{-19} \, \rm eV \lesssim m_s \lesssim 10^{-18} \, \rm eV$, our Markov Chain Monte Carlo analysis shows no evidence for such a cloud, placing an upper bound $M_{\rm cloud} \lesssim 0.1\% M_{\bullet}$ at $3\sigma$ confidence level.
We show that dynamical friction exerted by the medium on S2 motion plays no role in the analysis performed in this and previous works, and can be neglected thus.
\end{abstract}

\begin{keywords}
black holes physics -- dark matter -- gravitation -- celestial mechanics -- Galaxy: centre
\end{keywords}



\section{Introduction}
Since the star S2 has been discovered orbiting the Galactic Center (GC) \citep{2002Natur.419..694S, Ghez_2003, 2009ApJ...692.1075G, 2017ApJ...837...30G}, its orbital motion has been largely and extensively used to constrain the properties of the supermassive black hole (SMBH) Sagittarius A$^*$ (SgrA$^*$) and the environment around it. S2 is part of the so-called S-cluster, which currently counts up to tens of detected stars \citep{Sabha:2012vc, 2017ApJ...847..120H, 2022A&A...657A..82G}. 

The astrometric and spectroscopic data collected by two independent groups showed that the dynamics of S-stars is entirely dominated by the presence of a compact source with $M_{\bullet} \sim 4.3 \cdot 10^6 \, M_{\odot}$ at a distance of $R_0 \sim 8.3 \, \rm kpc$. There is overwhelming evidence that the compact source is a SMBH \citep{2002Natur.419..694S, 2008ApJ...689.1044G,  Genzel:2010zy, Genzel:2021, 2019A&A...625L..10G, GRAVITY:2021xju}. 
Very strong arguments that the central dark mass is indeed an SMBH come from the measurement of the Schwarzschild precession in the orbit of S2 \citep{GRAVITY:2020gka}, from the observations of near-IR flares in correspondence with the innermost circular orbit of the SMBH \citep{2018A&A...618L..10G, GRAVITY:2023avo} and by the image released by the Event Horizon Telescope collaboration, which is compatible with the expected image of a Kerr BH \citep{EventHorizonTelescope:2022wkp}. 

The physics of horizons is so puzzling that any further evidence for their existence is welcome and provides important information on the scales at which new physics sets in. Currently, it is challenging to use orbits of S-stars around the GC to test the nature of the compact source itself and to distinguish it from other possible models, such as boson stars, dark matter (DM) cores or wormholes, which have similar features to BHs \citep{Amaro-Seoane:2010pks, Grould_2017, Boshkayev:2018sbj, DellaMonica:2021fdr, deLaurentis:2022oqa}.
Note, however, that the optical appearance of hot spots (or stars) close to the accretion zone of SgrA$^*$, may differ significantly should an horizon be absent~\citep{Rosa:2022toh}.

Equally important is the nature of the environment around SMBHs, in particular around SgrA$^*$. Dark matter (DM) is expected to cluster at the center of galaxies leading to ``overdensities''\citep{Gondolo:1999ef, Sadeghian:2013laa}, which might leave an imprint in the motion of stars. S-stars are currently the main observational tool we have to look into this inner region of our Galaxy and thus they must be exploited to gain as much information as possible from their motion. For this and other reasons, the possibility of an extended mass distribution around SgrA$^*$ have been studied~\citep{Lacroix:2018zmg, Bar:2019pnz, Heissel:2021pcw,GRAVITY:2021xju, GRAVITY:2023cjt}. Specifically, \cite{GRAVITY:2021xju} derived an upper limit of $\delta M \sim 4000 \, M_{\odot} \sim 0.1\% M_{\bullet}$ for a density distribution described by a Plummer profile with length-scale $a_0 = 0.3''$. 

A special, and interesting, model for dark matter concerns ultralight bosons. These arise in a variety of scenarios, for instance the ``string axiverse" \citep{Arvanitaki:2009fg, Arvanitaki:2010sy, Marsh:2015xka} or as a hidden U(1) gauge boson, a generic feature of extensions of the Standard Model \citep{ Goodsell:2009xc, Jaeckel:2010ni}. In fact, such fields can exist and grow even if they are only a minute component of DM, as they are amplified via a mechanism known as BH superradiance~\citep{Brito:2015oca}. In this process, the light boson extracts rotational energy away from the spinning BH, depositing it in a ``bosonic cloud'', which can acquire a sizeable fraction of the BH mass. For a fundamental boson of mass $m_s$ the key parameter controlling the superradiant growth and energy extraction is the mass coupling $\alpha=M_{\bullet} m_s$.

In a recent work \citep{GRAVITY:2023cjt}, we investigated the possibility that a massive scalar field clusters around SgA$^*$ in the form of a cloud \citep{GRAVITY:2019tuf}. We showed that for the range of (dimensionless) mass couplings, $0.01 \lesssim \alpha \lesssim 0.045$ (which corresponds to a mass of the scalar field of $6 \cdot 10^{-19}\, \rm eV \lesssim m_s \lesssim 3 \cdot 10^{-18} \, \rm eV$) we are able to constrain the mass of the cloud to be $M_{\rm cloud} \lesssim 0.1 \% \, M_{\bullet}$, recovering the upper bound found in \cite{GRAVITY:2021xju}.

Here, we focus on a similar system: a massive vector cloud. 
As scalar fields, massive vector fields can form bound states around Kerr BHs, giving rise to stationary clouds. 
At the linear level and using the small coupling approximation, it has been shown that the superradiant instability is triggered on a timescale $\tau_I \propto \alpha^{-7}$ for vector clouds when compared to the scalar case of $\tau_I \propto \alpha^{-9}$ \citep{Pani:2012bp,Brito:2015oca,Cardoso:2018tly,2017JHEP...05..052E}. Hence vector clouds grow much faster than their scalar counterparts and the field's mass $m_s$ needed to make them grow in a timescale smaller than the cosmic age is much smaller, making them more likely to be observed. 

In this work we will use the astrometric and spectroscopic data of star S$2$ collected at the Very Large Telescope (VLT) to constrain the fractional mass of a possible vector cloud around SgrA$^*$. 

We will use units where $\hbar = c = G = 1$, unless otherwise stated.
\section{Setup}
In this work, we consider a massive vector field $A_{\mu}$ described by the Lagrangian 
\begin{equation}
    \mathcal{L} = - \frac{1}{4} F_{\mu \nu} F^{\mu \nu} - \frac{1}{2} \mu^2 A_{\mu} A^{\mu}
\end{equation}
and $A^{\mu}$ satisfies the Proca equation of motion $D_{\mu} F^{\mu \nu} = \mu^2 A^{\nu}$. If the Compton wavelength of the vector field is much larger than the Schwarzschild radius $r_g = M_{\bullet}$, the bound states of the field oscillate with frequency $\omega_f \simeq \mu$ and can be written as \citep{Baryakhtar:2017ngi}
\begin{equation}
    A^{\mu}(t, x) = \frac{1}{\sqrt{2 \mu}} \left(\Psi^{\mu}(x) e^{- i \omega_f t} + \rm c.c. \right)\, .
\end{equation}
In the limit $r \gg r_g$, the Proca equation becomes a Schr\"oedinger-like equation, and the $\Psi_0$ component can be expressed in terms of $\Psi_i$. Since the radial part of the potential is spherically symmetric, $\Psi_i$ can be decomposed as
\begin{equation}
    \Psi_i = R^{n \ell}(r) Y_i^{\ell, jm} (\theta, \phi) \, ,
\end{equation}
where the $ Y_i^{\ell, jm} (\theta, \phi)$ are the so-called pure-orbital vector spherical harmonics \citep{RevModPhys.52.299, Santos:2020pmh}. 

The fundamental mode of the field, which is also the mode that grows fastest due to superradiant mechanisms \citep{Baryakhtar:2017ngi} is 
given by 
$\ell = 0$, $m = j = 1$ and $n = 0$. At leading order in $\alpha$ we can neglect $A_0$ and consider only the spatial components of the field, which can be written as \citep{Chen:2022kzv}
\begin{equation}
A_i^{1011} = \Psi_0 e^{- \frac{\alpha^2 r}{M_{\bullet}}} \left(\cos(\mu t), \sin(\mu t), 0 \right)\, .
\label{vector_profile}
\end{equation}
From this profile, we can compute the energy-momentum tensor \citep{Herdeiro:2016tmi} and take the Newtonian limit, i.e. neglecting all the spatial derivatives and assuming a real field, obtaining:
\begin{equation}
    \rho = \frac{\Psi_0^2 \alpha^2}{M_{\bullet}^2} e^{-\frac{2 \alpha^2 r}{M_{\bullet}}} \, ,
    \label{density}
\end{equation}
which coincides with the expression in \cite{Chen:2022kzv}. 

As done in \cite{GRAVITY:2023cjt}, we can integrate the energy density in Eq.~\eqref{density} to relate the amplitude of the field $\Psi_0$ with the mass of the vector cloud:
\begin{equation}
    M_{\rm cloud} = \frac{\pi \Psi_0^2 M_{\bullet}}{\alpha^4}.
    \label{mcloud}
\end{equation}
From the energy density in Eq.~\eqref{density} we can get the potential generated by the cloud solving Poisson's equation: $\nabla^2 U_V = 4 \pi \rho$ and using the spherical harmonic decomposition of \cite{PoissonWill2012} to get:
\beq
U_V = \frac{\Lambda}{r} \left(M_{\bullet} - e^{- 2 r \alpha^2/M_{\bullet}} \left(M_{\bullet} + r \alpha^2 \right) \right)
\eeq
where we have defined $\Lambda = M_{\rm cloud}/M_{\bullet}$. 
\subsection{Effects of the cloud on S2 orbit with osculating elements}
\label{subsec:osculating_elements}
We start our analysis of the effects of vector cloud on S2 motion using the method of osculating elements that can be found in \cite{PoissonWill2012}. The basic idea is to treat the effect of the vector cloud as a perturbation of the Newtonian acceleration, assuming that the Keplerian description of the orbit is still approximately true. In this way, we are able to express the equations of motion in terms of the Keplerian elements $(e, a, i, \omega, \Omega, \mathcal{M}_0)$ (eccentricity, semi-major axis, inclination, argument of the periastron, longitude of the ascending node and mean anomaly at epoch, respectively), which would be constant in a pure Newtonian setup, and see how the perturbing force modifies them. In order to do so, we introduce a vectorial basis adapted to the orbital motion of the binary system BH-S2: $\left(\mathbf{n}, \boldsymbol{\lambda}, \mathbf{e}_z\right)$, where $\mathbf{n} = \mathbf{r}/r$, $\mathbf{e}_z = \mathbf{h}/h$ with $\mathbf{h} := \mathbf{r} \times \mathbf{v} $ and $\boldsymbol{\lambda}$ is orthogonal to both $\mathbf{n}$ and $\mathbf{e}_z$.  We also assume that the mass of the star is negligible compared to the BH mass $M_{\bullet}$. 

The perturbing force can be decomposed as:
\beq
\mathbf{f} = \mathcal{R} \mathbf{n} + \mathcal{S} \boldsymbol{\lambda} + \mathcal{W} \mathbf{e_z}
\label{perturbing_force}
\eeq
The variation of the orbital elements in terms of the perturbing force components is given in \cite{Kopeikin:2011mwv, PoissonWill2012} and we report it for completeness in Appendix \ref{app:orbital_elements}. 

 Once the variation in time of the orbital elements is known, one can compute the secular change of the orbital element $\mu^{a}$ over a complete orbit using:
\beq
\Delta \mu^a = \int_0^{2\pi} \frac{d \mu^a}{d\phi} d\phi \, ,
\eeq
where 
\beq
\frac{d \mu^a}{d\phi} = \frac{d \mu^a}{dt} \frac{dt}{d \phi}
\eeq
and  
\beq
\frac{d\phi}{dt} = \sqrt{\frac{M_{\bullet}}{a^3(1-e^2)^3}} (1 + e \cos \phi)^2 \, .
\eeq

\subsubsection{Effect of the vector cloud alone}

Due to the spherical symmetry of the energy distribution in Eq.~\eqref{density}, the only non-zero component of $\mathbf{f}_V$ is the radial one:
\beq
\mathcal{R}_V = \frac{\Lambda}{M_{\bullet} r^2} \left[- M_{\bullet}^2 + e^{-2 r \alpha^2/M_{\bullet}} \left(M_{\bullet}^2 + 2 M_{\bullet} r \alpha^2 + 2 r^2 \alpha^4\right) \right]
\label{R_vector}
\eeq
while $\mathcal{S}_V = \mathcal{W}_V = 0$. 
\subsubsection{Inclusion of the 1PN correction}
\label{subsec:schw_precession}

Since the Schwarzschild precession has been detected at $8 \sigma$ confidence level by the GRAVITY collaboration \citep{GRAVITY:2020gka, GRAVITY:2021xju}, it is interesting to see how the previous results change if we include the first Post Newtonian (PN) correction to the equations of motion. 

This corresponds to having a total acceleration
\beq
\mathbf{a} = - \frac{M_{\bullet} \mathbf{r}}{r^3} + \mathbf{a}_V + \mathbf{a}_{1\rm PN} \, , 
\label{total_a}
\eeq
where
\beq
\boldsymbol{a}_{1 \rm PN} = \frac{ M_{\bullet}}{r^2} \left[\left(\frac{4M_{\bullet}}{r} - v^2\right) \frac{\boldsymbol{r}}{r} + 4 \dot{r}\boldsymbol{v}  \right] \, ,
\label{1pn}
\eeq 
with $\boldsymbol{r} = r \hat{r}$, $\boldsymbol{v} = \left(\dot{r} \hat{r}, r \dot{\theta}\hat{\theta}, r \dot{\phi} \sin \theta \hat{\phi} \right)$ and $ v = |\boldsymbol{v}|$. 

The decomposition of the acceleration in Eq.~\eqref{1pn} into the basis $\left(\mathbf{n}, \boldsymbol{\lambda}, \mathbf{e}_z\right)$ has been done in \cite{PoissonWill2012} and here we report the result:
\beq
\mathcal{R}_{1\rm PN} = \frac{M_{\bullet}}{r^2} \left(4 \dot{r}^2 - v^2 + 4 \frac{M_{\bullet}}{r} \right) \, ,
\eeq
\beq
\mathcal{S}_{1\rm PN} = \frac{M_{\bullet}}{r^2} \left(4 \dot{r} r \dot{\phi} \right)\, , 
\eeq
and $\mathcal{W}_{1\rm PN} = 0$. In order to express everything in terms of the orbital elements, we need to use the expressions for $r$, $\dot{r}$ and $\dot{\phi}$ reported in Sec.10.1.3 of \cite{PoissonWill2012}. 

In this second case we set $\Lambda = 10^{-3}$, which corresponds to the current upper limit obtained by the GRAVITY collaboration for the fractional mass of an extended mass distribution around SgrA$^*$ \citep{GRAVITY:2021xju, GRAVITY:2023cjt}.

\subsection{Data}

The set of available data $D$ is the same as in \cite{GRAVITY:2023cjt}. 

\subsection{Fitting approach}
The next step is to obtain a best-fit value for the fractional mass $\Lambda$ for different coupling $\alpha$ values. The procedure followed in this work is exactly the same as the one reported in \cite{GRAVITY:2023cjt}. 
Specifically, we solve the equations of motion in Eq.~\eqref{total_a} using the initial conditions reported in Appendix~\ref{app:init_conditions}. The solutions of this set of equations are given in the BH reference frame and must be projected into the observer reference frame using the three Euler angles $\Omega, i, \omega$. 

Following \citet{Grould:2017bsw} we can define a new reference frame $\{x', y', z_{\rm obs}\}$ such that $x' = \rm DEC$, $y' = \rm R.A.$ are the collected astrometric data, $z_{\rm obs}$ points towards the BH and $v_{z_{\rm obs}}$ corresponds to the radial velocity (see Appendix~\ref{app:coord_transf} for details about how to perform the rotation of the reference frame). 

Moreover, it is true that S2 motion happens mostly in a Newtonian regime, i.e. with $v \ll 1$, but near the periastron, it reaches a total space velocity $v \sim 10^{-2}$. In this region, relativistic effects become important and can not be neglected. For this reason, we correct the radial velocity coming from Eq.~\eqref{total_a}, including both the relativistic Doppler shift and the gravitational redshift \citep{GRAVITY:2018ofz}. 

Finally, we also consider the so-called R{\o}mer delay, which is the difference between the observational dates and the actual emission dates of the signal due to the finite speed of light. Details about how to include R{\o}mer delay and relativistic effects are reported in Appendix~\ref{app:relativistic_effects}. 

For any given value of $\alpha$, we fit for the following set of parameters,
\begin{equation}
    \Theta_i = \{e, a, \Omega, i, \omega, t_p, R_0, M_{\bullet}, x_0, y_0, v_{x_0}, v_{y_0}, v_{z_0}, \Lambda \}\,.
    \label{emcee_parameters}
\end{equation}

The additional parameters $\{x_0, y_0,v_{x_0}, v_{y_0}, v_{z_0} \}$ characterise the NACO/SINFONI data reference frame with respect to Sgr~A* \citep{2015MNRAS.453.3234P}. We refer the reader to Appendix~\ref{app:mcmc_details} for more details about the MCMC implementation.
\section{Results}
\subsection{Variation of the orbital elements}
\label{variation_elements}
\begin{figure*}
\centering
\includegraphics[width=0.8\textwidth]
{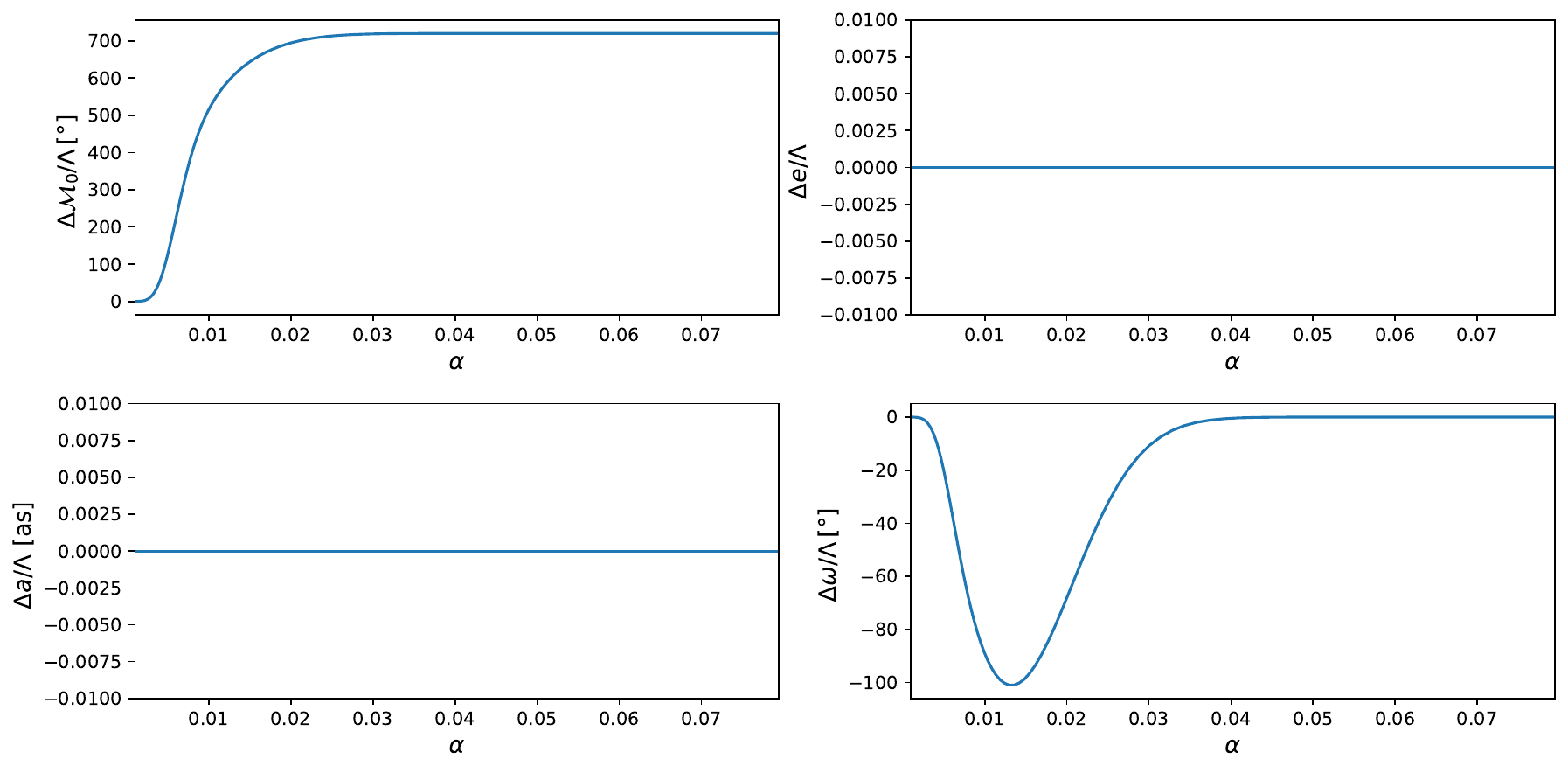}
\caption{Variation of the orbital elements $\Delta \mu^{a}/\Lambda$ over an entire orbit for different values of the coupling constant $\alpha$ when only the vector cloud is present. The maximum variation in $\Delta \omega/\Lambda$ is roughly found in the range $0.003 \lesssim \alpha \lesssim 0.03$.}
\label{fig:var_elements_cloud}
\end{figure*}

In Figure~\ref{fig:var_elements_cloud} we show the variation of the orbital elements $\Delta \mu^{a}/\Lambda$ due to the presence of the vector cloud for different values of the coupling $\alpha$, as described in Sec.~\ref{subsec:osculating_elements}. The secular change is negligible for both the eccentricity $e$ and the semi-major axis $a$. 

The change in the mean anomaly at epoch $\mathcal{M}_0$ is instead proportional to $\alpha$, increasing monotonically. $\mathcal{M}_0$ is directly related to the time of pericenter passage $t_p$: a larger mean anomaly at the epoch corresponds to a later pericenter passage. 

The only meaningful change in the orbital elements is found in  $\Delta \omega$, which quantifies the precession effect on the orbit, with $\omega$ the argument of pericenter. First of all, we observe that $\Delta \omega < 0$ always. This is a consequence of the fact that the presence of an extended mass within the orbit of S$2$ would produce a retrograde precession of the orbit \citep{Heissel:2021pcw}. 

Unsurprisingly, its maximum variation is found in the range 
\beq
0.003 \lesssim \alpha \lesssim 0.03 \, .
\label{range_alpha}
\eeq

Indeed, as in the case of scalar clouds \citep{GRAVITY:2023cjt}, this behaviour is expected if we compute the effective peak position of the energy distribution in Eq.~\eqref{density}, 
\beq
R_{\rm peak} = \frac{\int_0^{\infty} \rho r dr}{\int_0^{\infty} \rho dr} = \frac{M_{\bullet}}{2 \alpha^2} \, ,
\label{r_peak}
\eeq
which, for the values of $\alpha$ reported in Eq.~\eqref{range_alpha}, corresponds to $5 \cdot 10^2 \, M_{\bullet} \lesssim R_{\rm peak} \lesssim 5 \cdot 10^4 \, M_{\bullet}$, i.e. it roughly matches the orbital range of S$2$ ($3 \cdot 10^3 \, M_{\bullet} \lesssim r_{s2} \lesssim 5 \cdot 10^4 \, M_{\bullet}$). This result shows that the maximum variation in $\omega$ is found when the star crosses regions of higher (vector) density, while its orbit remains basically unaffected if the cloud is located away from its apoastron or too close to the central BH mass. 

\begin{figure*}[h]
\centering
\includegraphics[width=0.8\textwidth]
{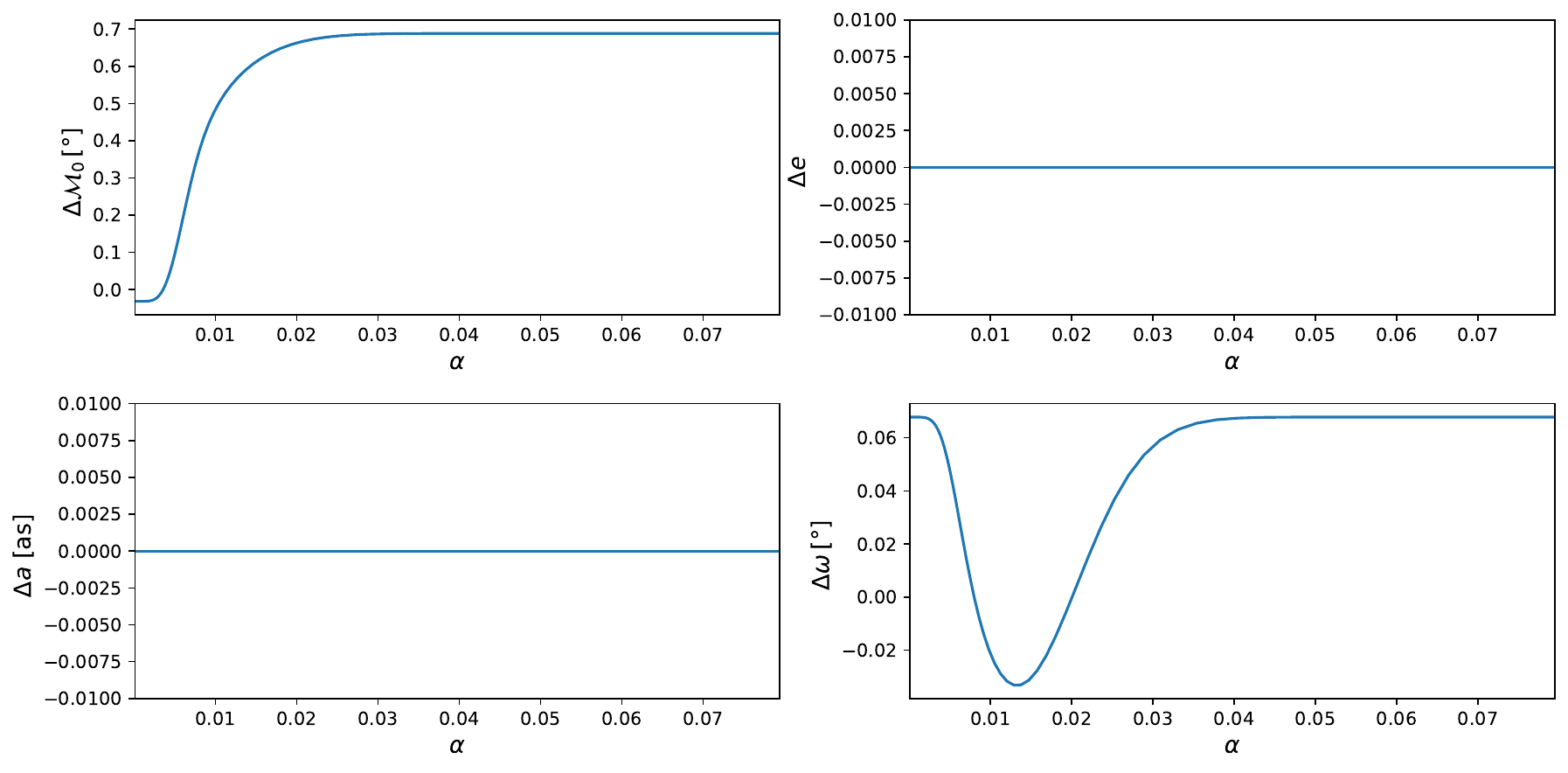}
\caption{Variation of the orbital elements $\Delta \mu^{a}$ over an entire orbit for different values of the coupling constant $\alpha$ when one includes the Schwarzschild precession in the equation for the osculating elements. Here $\Lambda = 10^{-3}$. The maximum variation is still found in $0.003 \lesssim \alpha \lesssim 0.03$.}
\label{fig:var_elements_cloud_1PN}
\end{figure*}

In Figure~\ref{fig:var_elements_cloud_1PN} we show the variation of the orbital elements when the 1PN correction is included in the equations of motion, as described in Sec.~\ref{subsec:schw_precession}. Opposite to the previous case, here, the variation of the argument of the pericenter $\Delta \omega$ can be either positive or negative, according to the value of $\alpha$. Indeed now the retrograde precession induced by the vector cloud is compensated by the (prograde) Schwarzschild precession due to the 1PN correction in the equations of motion, and its maximum value corresponds to $\Delta \omega \simeq -1.8'$, which is smaller than the previous case with $\Lambda = 10^{-3}$ ($\Delta \omega \simeq -6'$). 

\subsection{Limit on the fractional mass $\Lambda$}
\label{mcmc_analysis}

\begin{figure*}
\centering
\includegraphics[width=0.6\textwidth]
{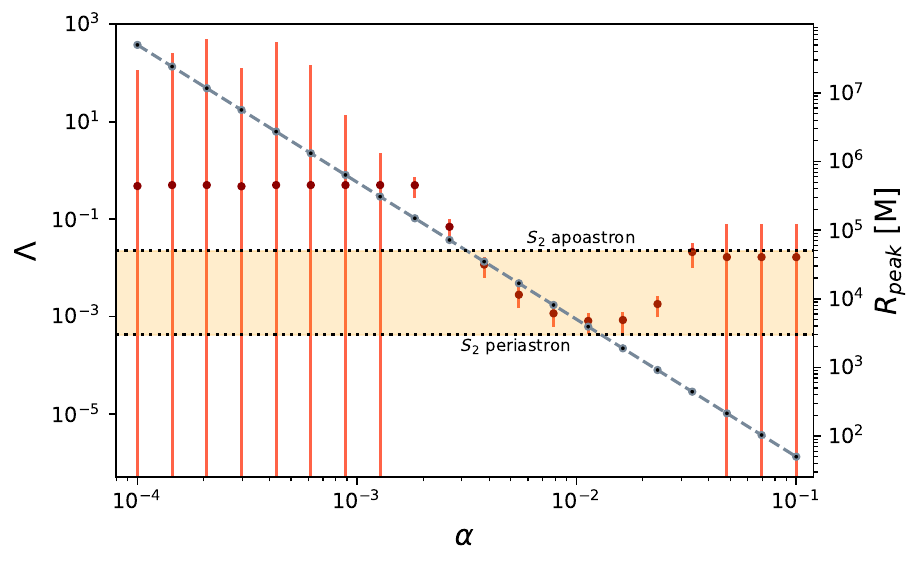}
\caption{Best-fit values for $\Lambda$ and relative $1\sigma$ uncertainties as function of the coupling $\alpha$ obtained minimizing the $\chi^2$. The grey dashed line represents the effective peak position of the vector cloud given by Eq.~\eqref{r_peak}, while the orange band gives the orbital range of S2.}
\label{fig:chi_squared_analysis}
\end{figure*}

Before running the MCMC algorithm we determine the initial guesses for the parameters listed in Eq.~\eqref{emcee_parameters}. We performed a simple $\chi^2$ minimization using the Python package \textbf{lmfit.minimize} \citep{lmfit2014} with Levenberg-Marquardt method. 
In Figure~\ref{fig:chi_squared_analysis} we report the best-fit values of $\Lambda$ with relative $1\sigma$ uncertainties, and we compare the range of $\alpha$ with the effective peak position of the cloud in Eq.~\eqref{r_peak}. The smallest uncertainties for $\Lambda$ are found roughly in the range of Eq.~\eqref{range_alpha}, which is slightly different from the scalar cloud case \citep{GRAVITY:2023cjt} and in agreement with the orbital variation reported in Figure~\ref{fig:var_elements_cloud_1PN}. 
\begin{figure*}
\centering
\includegraphics[width=\textwidth]
{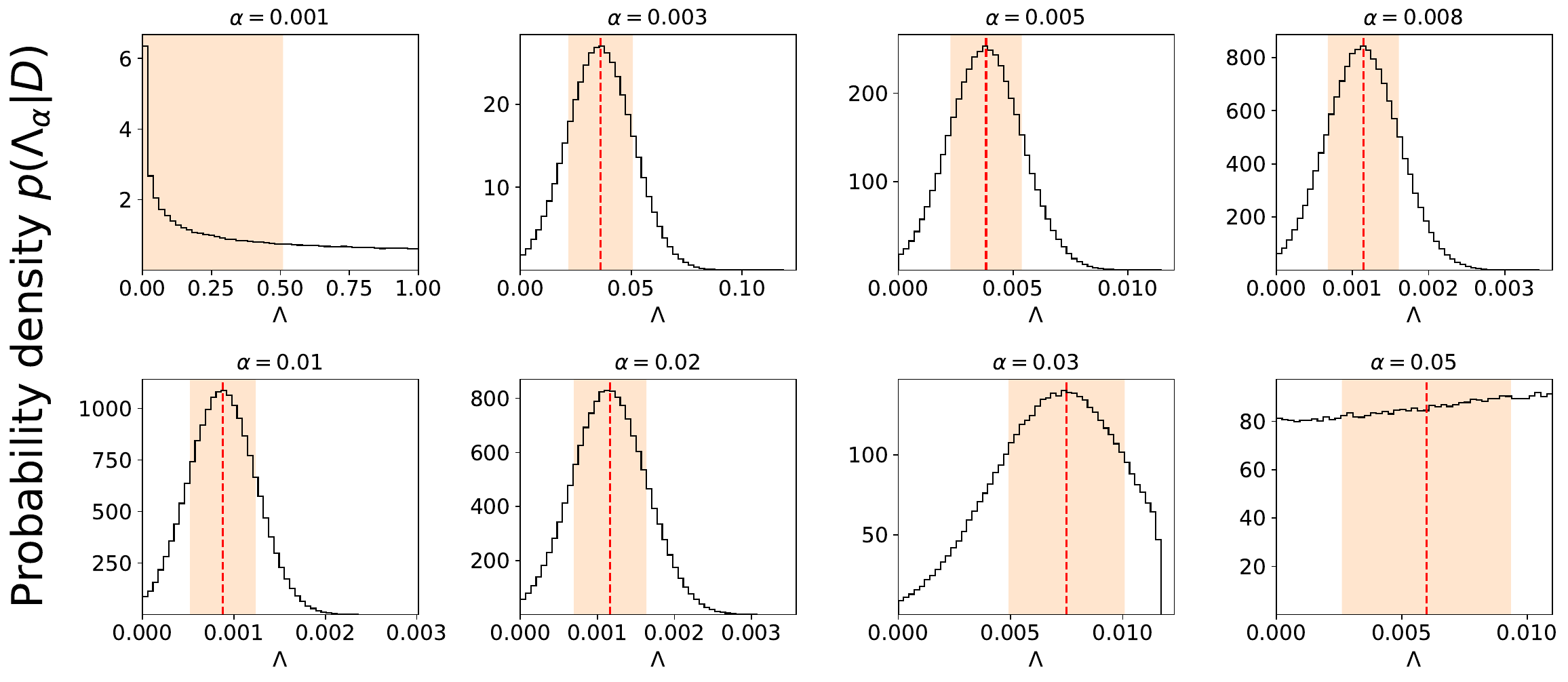}
\caption{Posterior probability densities $p(\Lambda_{\alpha}|D)$ for different values of $\alpha$. Red dashed lines represent the mean value of the distributions (which coincides with the MLE $\hat{\Lambda}$), while orange bands correspond to $1 \sigma$ confidence level, such that $\approx 68\%$ of $p(\Lambda_{\alpha}|D)$ lies in that region.}
\label{fig:lambdas}
\end{figure*}

After performing the MCMC analysis, we look for the maximum likelihood estimator (MLE) $\hat{\Lambda}$, which in this case corresponds to the value that maximises the posterior density distribution reported in Figure~\ref{fig:lambdas}, as a consequence of using flat priors and a Gaussian likelihood.

\begin{table}
\caption{Maximum Likelihood Estimator $\hat{\Lambda}$ with associated $1 \sigma$ error and Bayes factors $\log_{10} K$ for different values of $\alpha$. The measurements for each $\alpha$ are not independent (the same orbit was used to derive them) and therefore cannot be combined to derive a more stringent upper limit.
For non-normal distributions we report $\Lambda_1$ and $\Lambda_2$ defined such that $P(\Lambda_{\alpha}<\Lambda_1|D) \approx 68\%$ and $P(\Lambda_{\alpha}<\Lambda_2|D) \approx 99\%$ of $P(\Lambda_{\alpha}|D)$.} 
\begin{tabular}{lll}
\hline
$\alpha$ & $\hat{\Lambda}$ & $\log_{10} K$  \\
\hline
    $ 0.001$ & $\lesssim (0.51, 0.98)$ & -0.45  \\ \hline
    $ 0.003$ & $0.03596 \pm 0.01477$ & -2.09  \\ \hline
    $ 0.005$ & $0.00379 \pm 0.00157$ & -3.11  \\ \hline
    $ 0.008$ & $0.00114 \pm 0.00047$ & 1.62 \\ \hline
    $ 0.01$ & $0.00088 \pm 0.00036$ & 1.42 \\ \hline
    $ 0.02$ &  $0.00116 \pm 0.00047$ & 1.69 \\ \hline  
    $ 0.03$ & $0.00688 \pm 0.00263$ & -2.55 \\ \hline 
    $ 0.04$ & $0.00617 \pm 0.00337$ & -4.77\\  \hline 
    $ 0.05$ & $0.00592 \pm 0.00339$ & -4.96 \\ 
    \hline
\end{tabular}
\label{table:best_fit_lambda}
\end{table}

In Table~\ref{table:best_fit_lambda} we report the values of $\hat{\Lambda}$ with relative $1\sigma$ uncertainties together with the value of the Bayes factor $\log K$. The latter is obtained computing the marginal likelihoods by making use of the Python package \textbf{MCEvidence} developed in \cite{Heavens:2017afc} and it is defined as $K = P(D|M_{\alpha})/P(D|M_0)$, where $M_{\alpha}$ represents the BH plus vector cloud model while $M_0$ corresponds to the non perturbative one.


When the posterior distribution is found to be non-normal and peaked at zero, we estimated the $1\sigma \, (3\sigma)$ confidence interval looking for that value of $\Lambda$ such that roughly the $68\% \, (99\%)$ of $p(\Lambda|D)$ lies below that value. 
When $\alpha \gtrsim 0.3$, the distribution of $\Lambda$ start to be flat, with a sudden drop around $\Lambda \simeq 10^{-2}$. One can show that for flat distributions in an interval $[a, b]$, 
the mean is given by $(a-b)/2$ while the variance is $(b-a)^2/12$ \citep{bailer-jones_2017}. We report those values in Table~\ref{table:best_fit_lambda}. However, what is important to notice in these cases is that for $\alpha \gtrsim 0.03$ ($R_{\rm peak} \lesssim 550 \, M_{\bullet}$), it is not possible to determine a unique value for $\Lambda$ that best fits the data, confirming the expectation from the $\chi^2$ minimisation. 

When $\alpha$ is in the range of Eq.~\eqref{range_alpha} the posterior distributions of $\Lambda$ are Gaussian whose means and standard deviations are reported in Table~\ref{table:best_fit_lambda}. For all cases considered in this range, $\hat{\Lambda} \sim 10^{-3}$ with $1 \sigma$ uncertainties roughly of the same order of magnitude. This makes all the $\hat{\Lambda}$ values derived from the MCMC analysis compatible with zero within the $3\sigma$ confidence level. In addition to this, the associated Bayes factors always have $\log K < 2$. This result, according to the literature \citep{Kass:1995loi}, shows no statistical evidence in favour of the BH plus vector cloud model with respect to the non-perturbative case where no cloud is present. Hence we derive an upper limit of $\Lambda \lesssim 10^{-3}$ at $3\sigma$ confidence level.

This upper bound imposes a limit on the superradiant growth, that in general would lead to transfer up to $\sim \mathcal{O}(10)\%$ of the BH mass into the vector cloud \citep{Brito:2014wla, East:2017ovw, Herdeiro:2021znw}. Here we showed that for a field's effective mass of $m_s \sim 10^{-19}-10^{-18} \, \rm eV$, the mass of the cloud around SgrA$^*$ can not exceed the limit $M_{\rm cloud} \lesssim 0.1\% M_{\bullet}$.
For a BH spinning with $a/M \sim 0.5$ (an indicative value), the growth timescale of the cloud can vary between $10^5 - 10^{10}  \, \rm yrs$, exact values depend on the effective mass $m_s$. This estimate is below the age of the Universe ($t_{\rm age} \sim 10^{10} \, \rm yrs$), making the superradiant process and our constraints relevant. In Appendix \ref{app:corner_plot} we report the corner plots of two illustrative cases ($\alpha = 0.01$, $\alpha = 0.001$) to show the correlations between parameters.

\subsection{Inclusion of environmental effects}
\label{environmental_effect}
All the above results are obtained neglecting the backreaction effects of the matter on the motion of S2. Indeed, the presence of a matter distribution induces a gravitational drag force on the body moving in it, with the consequence that part of the material is dragged along the motion producing dynamical friction force on the main body \citep{1983mtbh.book.....C, Ostriker_1999}. It has been shown that dynamical friction induced by ultralight bosons may play a significant role in the strong regime \citep{Traykova:2021dua, Vicente:2022ivh}. Here we investigated whether dynamical friction affects S2 motion too. 

In a Newtonian setup, including the dynamical friction force means adding the following two components to the equations of motion \citep{Macedo:2013qea}:
\beq
\begin{split}
    & F_{\rm DF, r} = F_{\rm DF} \frac{\dot{r}}{v} \\
    & F_{\rm DF, \phi} = F_{\rm DF} \frac{r \dot{\phi}}{v}
\end{split}
\label{DF_force}
\eeq
where $v^2 = \dot{r}^2 + r^2 \dot{\phi}^2$, since we have assumed that the motion of S2 happens on the equatorial plane ($\theta = \pi/2$) of the central SMBH. 

The term $F_{\rm DF}$ has been derived in \cite{Ostriker_1999} for a perturber in linear motion and it reads:
\beq
F_{\rm DF} = - \frac{4 \pi \mu_s^2 \rho}{v^2} I_v
\label{DF_linear}
\eeq
with
\beq 
I_v = \begin{cases} 
        \frac{1}{2} \log\left(\frac{1 + v/c_s}{1 - v/c_s}\right) - \frac{v}{c_s} \, , & v < c_s \\
        \frac{1}{2} \log\left(1 - \frac{c_s^2}{v^2}\right) + \log \left(\frac{v t}{r_{\rm min}}\right)\, , & v > c_s ,
        \end{cases}
\label{Iv}
\eeq
where $\rho$ is the density of the matter distribution in Eq.~\eqref{density}, $\mu_s$ is the mass of the star S2 that we take to be $\mu_s = 14 \, M_{\odot}$ and $c_s$ is the speed of sound in the medium which constitutes
the environment. \cite{Kim:2007zb} showed that Eq.~\eqref{DF_linear} correctly reproduces the results obtained for circular orbits if one substitutes $ vt \rightarrow 2 r(t)$. 

Despite the orbit of S2 is far from being circular, we are going to use Eqs.~\eqref{DF_force} in a first approximation. 

We tested four different values of the speed of sound $c_s$ for both the supersonic ($c_s = 10^{-6}$, $c_s = 10^{-3}$) and the subsonic ($c_s = 0.1$, $c_s = 0.03$) regimes, for different values of $\alpha$. We set $\Lambda = 10^{-3}$, since this corresponds to the maximum allowed value of the fractional mass, but results scale linearly with it.

We found that results are independent on $c_s$ and that the maximum difference in both the astrometry and the radial velocity with respect to the case where no dynamical friction is implemented is always negligible. 

\begin{figure*}
\centering
\includegraphics[width=\textwidth]
{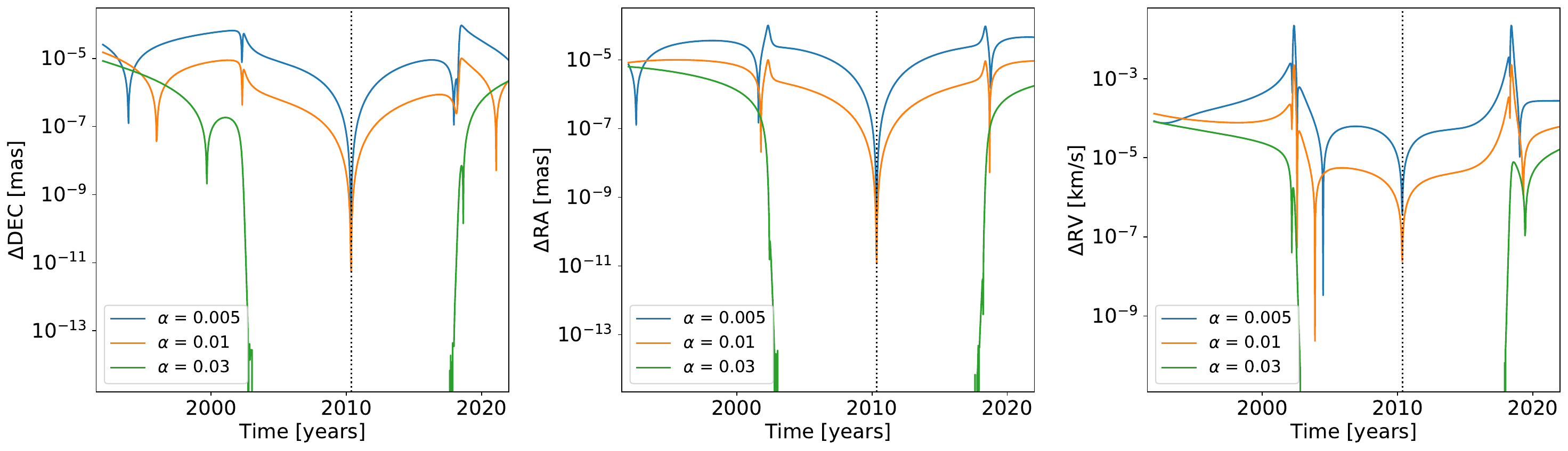}
\caption{Absolute difference in DEC, R.A. and radial velocity between the case where dynamical friction is implemented in the supersonic case with $c_s = 10^{-3}$ and the case where no dynamical friction is present. We set $\Lambda = 10^{-3}$, but results scale linearly with $\Lambda$. The difference is maximum around the periastron passages and minimum at the apoastron (black dotted line). Overall, they remain far below the current instrument threshold, whatever the value of $\alpha$.}
\label{fig:dynamical_friction}
\end{figure*}

In Figure~\ref{fig:dynamical_friction} we report the absolute difference in DEC, R.A. and radial velocity in the supersonic case with $c_s = 10^{-3}$. Overall, the effect of dynamical friction is at most $10^{-5} \, \rm mas$ in the astrometry and $\approx 10^{-3} \, \rm km/s$ in the radial velocity, and in both cases is reached around the periastron passages. Overall, it remains well below the current (and future) instrument precision and can be neglected. 

We performed the same analysis for the scalar cloud model implemented in \cite{GRAVITY:2023cjt} and the Plummer density profile tested in \cite{GRAVITY:2021xju} too. In both cases, we found similar results to Figure~\ref{fig:dynamical_friction} and hence we conclude that dynamical friction effects can be safely neglected.   

Along the same line, one can try to compute the effect that regular gas around SgrA$^*$ has on S$2$ orbit.
In \cite{Gillessen:2019}, the authors detected a drag force acting on the gas cloud G$2$ orbiting around SgrA$^*$ and they derived an estimate for the number density of the ambient. Here we used their same formulation for the drag force, meaning 
\beq
F_{\rm drag} = c_D r^{-\gamma} v^2 \mu_s \, ,
\label{drag_force}
\eeq
where $\gamma = 1$, $v$ is the relative velocity between the medium and the star, that, following \cite{Gillessen:2019}, is assumed to be equal to the velocity of the star itself and $c_D$ parametrizes the strength of the drag force and it is related to the normalized number density of the gas ambient. In \cite{Gillessen:2019} they derived $c_D \sim 10^{-3}$, which is the value used in this work as well. In this case no vector cloud is present ($\Lambda = 0$) and only the force contribution due to the presence of gas is considered. 

The maximum difference induced by the drag force exerted by the gas ambient on the astrometry and the radial velocity of S$2$ is of order $\sim 10^{-6} \, \rm mas$ and $\sim 10^{-3}\, \rm km/s$, respectively. Hence, also the contribution due to regular gas around SgrA$^*$ has a negligible effect on S$2$. We also note that the difference induced by the presence of gas is comparable with the effect produced by dynamical friction. Hence, even with the development of future instruments and the advent of GRAVITY+, it will still be hard to disentangle the two effects. 

\section{Conclusions}
In this paper we investigated the possibility that a vector cloud of superradiant origin clusters around the SMBH SgrA$^*$, extending the analysis on scalar clouds performed in \cite{GRAVITY:2023cjt}. Specifically, we considered a massive vector field, which gives rise to a spherically symmetric cloud and in Sec.~\ref{variation_elements} we investigated the imprints of such a cloud in S2's orbital elements. The MCMC analysis in Sec.~\ref{mcmc_analysis} confirmed the current upper bound for the fractional mass of $\Lambda \lesssim 0.1\% M_{\bullet}$, recovering previous results on extended masses \citep{GRAVITY:2021xju, GRAVITY:2023cjt}. Despite the range of field's masses that can be tested with S$2$ motion is roughly the same in both the scalar and vector cloud case ($10^{-18}\, \rm eV \lesssim m_s \lesssim 10^{-19}\, eV$), in the latter those values can effectively engage a superradiant instability in a timescale shorter than the cosmic age. This strongly constrains the mass of a possible superradiant cloud at the GC, improving the theoretical bound that can lead to have masses up to two order of magnitude larger~\citep{Brito:2014wla, East:2017ovw, Herdeiro:2021znw}. 

Moreover, the effect of the environment on S2 orbit was also investigated for the first time. We considered both the dynamical friction exerted by the medium on the star, and the effect of ambient gas around SgrA$^*$. In both cases, the effect on the astrometry and the radial velocity are negligible. This analysis was also extended to the scalar cloud case considered in~\cite{GRAVITY:2023cjt} and to the Plummer profile of~\cite{GRAVITY:2021xju}, showing that even in those cases both effect can be neglected. However, since the difference in the astrometry and the radial velocity induced by those effects is of the same order of magnitude, it will be difficult to separate them even with the advent of future instrumentation.

\section*{Acknowledgements}
The authors would like to thank the anonymous referee and Jarle Brinchmann for their suggestions that improved our work. We are very grateful to our funding agencies (Max Plank Gesellschaft, European Research Council (ERC), Centre National de la Recherche Scientifique [PNCG, PNGRAM], Deutsche Forschungsgemeinschaft, Bundesministerium f\"ur Bildung und Forschung, Paris Observatory [CS, PhyFOG], Observatoire des Sciences de l'Univers de Grenoble, and the Funda\c c\~ao para a Ci\^encia e a Tecnologia), to European Southern Observatory and the Paranal staff, and to the many scientific and technical staff members in our institutions, who helped to make NACO, SINFONI, and GRAVITY a reality. 
This project has received funding from the European Union's Horizon 2020 research and innovation programme under the Marie Sklodowska-Curie grant agreement No 101007855.
We acknowledge the financial support provided by FCT/Portugal through grants 
2022.01324.PTDC, PTDC/FIS-AST/7002/2020, UIDB/00099/2020 and UIDB/04459/2020. We acknowledge the funds from the ``European Union NextGenerationEU/PRTR'', 
Programa de Planes Complementarios I+D+I (ref. ASFAE/2022/014). 

\section*{Data Availability}
Publicly available data for astrometry and radial velocity up to 2016.38 can be found in Table 5 the electronic version of \cite{2017ApJ...837...30G} at this link:  \url{https://iopscience.iop.org/article/10.3847/1538-4357/aa5c41/meta#apjaa5c41t5}.



\bibliographystyle{mnras}
\bibliography{biblio} 




\appendix

\section{Variation of the orbital elements}
\label{app:orbital_elements}
The variation of the orbital elements in terms of the perturbing force in Eq.~\eqref{perturbing_force} is given by
\beq
\frac{d a}{dt} = 2 \sqrt{\frac{a^3}{M_{\bullet}(1-e^2)}} \left[e \sin \phi \mathcal{R} + (1 + e \cos \phi) \mathcal{S} \right] \, ,
\eeq
\beq
\frac{de}{dt} = \sqrt{\frac{a(1-e^2)}{M_{\bullet}}} \left[ \sin \phi \mathcal{R} + \frac{2 \cos \phi + e(1 + \cos^2 \phi)}{1 + e \cos \phi} \mathcal{S} \right] \, ,
\eeq
\beq
\begin{split}
\frac{d \omega}{dt} & = \frac{1}{e}\sqrt{\frac{a(1-e^2)}{M_{\bullet}}} \left[-\cos \phi \mathcal{R} + \frac{1 + 2 e \cos \phi}{1 + e \cos \phi} \sin \phi \mathcal{S} \right. \\
& \left. - e \cot i \frac{\sin(\omega + \phi)}{1 + e \cos \phi} \mathcal{W} \right]\, ,
\end{split}
\eeq
\beq
\frac{d i}{dt} = \sqrt{\frac{a(1-e^2)}{M_{\bullet}}} \frac{\cos(\omega + \phi)}{1 + e \cos \phi} \mathcal{W} \, ,
\eeq
\beq
\sin i \frac{d \Omega}{dt} = \sqrt{\frac{a(1-e^2)}{M_{\bullet}}} \frac{\sin(\omega + \phi)}{1 + e \cos \phi} \mathcal{W} \, ,
\eeq
and 
\beq
\frac{d \mathcal{M}_0}{dt} = - \sqrt{1-e^2}\left(\frac{d \omega}{dt} + \cos i \frac{d\Omega}{dt} \right) - \sqrt{\frac{a}{M_{\bullet}}} \frac{2 (e^2 -1)}{(1 + e \cos \phi)} \mathcal{R}
\eeq
where we have used the substitution $r = a(1-e^2)/(1+ e \cos \phi)$.

\section{Initial conditions and Kepler equation}
\label{app:init_conditions}
Since we start our numerical integration at apoastron, the 6 initial conditions for the set of equations in Eqs.~\eqref{total_a} can be obtained from the analytical solution of the Keplerian two-body problem, namely
\beq
\begin{split}
    & r_0 = \frac{a(1 - e^2)}{1 + e \cos \phi_0}\, , \,\,\,\,\,\,\,\,\,\,\,\,\,\,\,\,\,\,\,\,\,\,\,\,\,\,\,\,\,\,\,\,\,\,\,\, \dot{r}_0 =  \frac{2 \pi e a \sin \mathcal{E}}{P(1 - e \cos \mathcal{E})} \, ,\\
    & \theta_0 = \frac{\pi}{2} \, , \,\,\,\,\,\,\,\,\,\,\,\,\,\,\,\,\,\,\,\,\,\,\,\,\,\,\,\,\,\,\,\,\,\,\,\,\,\,\,\,\,\,\,\,\,\,\,\,\,\,\,\,\,\,\,\,\,\,\,\,\, \dot{\theta} = 0 \, ,\\
    & \phi_0 = 2 \arctan\left(\sqrt{\frac{1 + e}{1 - e}} \tan \frac{\mathcal{E}}{2} \right) \, , \,\,\,\, \dot{\phi}_0 = \frac{2 \pi (1-e)}{P(e \cos \mathcal{E} - 1)^2} \sqrt{\frac{1 + e}{1 - e}}\, ,
\end{split}
\label{initial_cond}
\eeq
where $e, a, P$ are the eccentricity, the semi-major axis and the period of the orbit, respectively, while $\mathcal{E}$ is the eccentric anomaly evaluated from Kepler's equation: $\mathcal{E} - e \sin \mathcal{E} - \mathcal{M} = 0$, where $\mathcal{M} = \mathcal{M}_0 + n (t - t_p)$ is the mean anomaly, $n = \sqrt{M_{\bullet}/a^3}$ is the mean angular velocity and $t_p$ is the time of periastron passage. 

Kepler's equation is solved using a Python's root finder (\textbf{scipy.optimize.newton}) which implements a Newton-Raphson method. The latter solves the equation with a precision of $\mathcal{O}(10^{-16})$. 

\section{Coordinate transformation}
\label{app:coord_transf}

The transformation from the orbital reference frame to the observer reference frame can be achieved using the following conversion:
\beq
\begin{split}
& x' = A x_{\rm BH} + F y_{\rm BH} \,\,\,\,\,\,\,\,\,\,\,\,\,\,\,\,\,\,\,\,\, v_{x'} = A v_{x_{\rm BH}} + F v_{y_{\rm BH}} \\
& y' = B x_{\rm BH} + G y_{\rm BH}\,\,\,\,\,\,\,\,\,\,\,\,\,\,\,\,\,\,\,\,\, v_{y'} = B v_{x_{\rm BH}} + G v_{y_{\rm BH}} \\
& z_{\rm obs} = -(C x_{\rm BH} + H y_{\rm BH}) \,\,\,\,\,\,\,\,\,\,\,\, v_{z_{\rm obs}} = -(C v_{x_{\rm BH}} + H v_{y_{\rm BH}}) \, ,
\end{split}
\label{coord_obs}
\eeq 
where $A, B, C, F, G, H$ are the Thiele-Innes parameters \citep{Catanzarite:2010wa} defined as:
\begin{equation}
\begin{split}
& A = \cos \Omega \cos \omega -\sin \Omega \sin \omega \cos i \\
& B = \sin \Omega \cos \omega + \cos \Omega \sin \omega \cos i \\
& F = -\cos \Omega \sin \omega - \sin \Omega \cos \omega \cos i \\
& G = -\sin \Omega \sin \omega + \cos \Omega \cos \omega \cos i \\
& C = - \sin \omega \sin i \\
& H = -\cos \omega \sin i \,, 
\end{split}
\end{equation}
while the Cartesian coordinates $\{x_{\rm BH}, y_{\rm BH}, z_{\rm BH}\}$ and velocities $\{v_{x_{\rm BH}}, v_{y_{\rm BH}}, v_{z_{\rm BH}}\}$ are those obtained from the numerical integration. For a more detailed discussion about how the coordinate system $\{x', y', z_{\rm obs}\}$ and the above transformation are defined we refer the reader to Figure~1 and Appendix~B of \citet{Grould:2017bsw}. 

\section{Relativistic effects and R{\O}mer's delay}
\label{app:relativistic_effects}

As said in the main text, there are two main contributions that must be taken into consideration when S2 approaches the periastron: the relativistic Doppler shift and the gravitational redshift. Both of them induce a shift in the spectral lines of S2 that affects the radial velocity measurements. 
The former is given by
\beq 
1 + z_{D} = \frac{1 + v_{z_{\rm obs}}}{\sqrt{1- v^2}} \,,
\eeq 
while the gravitational redshift is defined as 
\beq 
1 + z_{\rm G} = \frac{1}{\sqrt{1 - 2 M/r_{\rm em}}} \,.
\eeq
The two shifts can be combined using Eq.~(D.13) of \citet{Grould:2017bsw} to obtain the total radial velocity 
\beq
V_R \approx \frac{1}{\sqrt{1 - \epsilon}} \cdot \frac{1 + v_{z_{\rm obs}}/\sqrt{1-\epsilon}}{\sqrt{1 - v^2/(1 - \epsilon)}} - 1 \,,
\eeq
where $\epsilon = 2M/r_{\rm em}$. 

In the total space velocity $v = |\textbf{v}|$ we must also add a correction due to the Solar System motion. We followed the most recent work of \citet{2020ApJ...892...39R} and take a proper motion of Sgr~A* of 
\beq
\begin{split}
& v_x^{\rm SSM} = -5.585 \, \rm mas/yr = 6.415 \cos(209.47^{\circ}) \, mas/yr \, ,\\
& v_y^{\rm SSM} = -3.156 \, \rm mas/yr = 6.415 \sin(209.47^{\circ}) \, mas/yr \, .
\end{split}
\eeq

The R{\o}mer's delay is instead included using the first order Taylor's expansion of the R{\o}mer's equation $t_{\rm obs} - t_{\rm em} - z_{\rm obs}(t_{\rm em}) = 0$, which reads:
\begin{equation}
    t_{\rm em} = t_{\rm obs} - \frac{z_{\rm obs}(t_{\rm obs})}{1 + v_{z_{\rm obs}}(t_{\rm obs})} \,.
    \label{t_em}
\end{equation}
\noindent The difference between the exact solution and the approximated one in Eq.~\eqref{t_em} is at most $\sim 4$ s over S2 orbit and therefore negligible. The R{\o}mer effect affects both the astrometry and the spectroscopy, with an impact of $\approx 450 \, \mu$as on the position and $\approx 50$ km/s at periastron for the radial velocity. Our results recover the previous estimates for this effect in \citet{Grould:2017bsw, GRAVITY:2018ofz}.

\section{MCMC details}
\label{app:mcmc_details}
We used a Gaussian log-likelihood given by
\begin{equation}
    \ln \mathcal{L} = \ln \mathcal{L}_{\rm pos} + \ln \mathcal{L}_{\rm vel}\,,
\end{equation}
where 
\begin{equation}
    \ln \mathcal{L}_{\rm pos} = - \sum_{i=1}^{N} \left[ \frac{ (\rm DEC_{i} - \rm DEC_{\rm model, i})^2}{\sigma_{\rm DEC_{i}}^2}  +  \frac{ (\rm R.A._{i} - \rm R.A._{\rm model, i})^2}{\sigma_{\rm R.A._{i}}^2} \right]\,,
\end{equation}

and 
\begin{equation}
    \ln \mathcal{L}_{\rm vel} = - \sum_{i=1}^{N} \frac{ (V_{R, i} - V_{\rm model, i})^2}{\sigma_{V_{R, i}}^2} \, .
\end{equation}
The priors we used are listed in Table~\ref{table:priors}. We used uniform priors for the physical parameters, i.e. we only imposed physically motivated bounds and Gaussian priors for the additional parameters describing NACO data, since the latter have been well constrained by previous work \citep{2015MNRAS.453.3234P} and are not expected to change.
\begin{table}
\caption{Uniform priors used in the MCMC analysis. Initial guesses $\Theta_i^0$ coincide with the best-fit parameters found by \textbf{minimize}.} 
\label{table:priors}
\begin{tabular}{lccc}
    \hline
    Parameter & $\Theta_i^0$ & Lower bound & Upper bound \\
    \hline
    $e$ & 0.88441 & 0.83 & 0.93 \\[2pt] 
    $a_{\rm sma}$ [as] & 0.12497 & 0.119 & 0.132 \\[2pt]  
    $i_{\rm orb} \, [^\circ] $  & $134.69241$ & 100 & $150$ \\ [2pt] 
    $\omega_{\rm orb} \, [^\circ]$ & $66.28411$ & 40 & $90$ \\ [2pt] 
    $\Omega_{\rm orb} \, [^\circ]$ & $228.19245$ & $200$ & $250$ \\ [2pt] 
    $t_p $ [yr] & 2018.37902 & 2018 & 2019 \\ [2pt] 
    $M_{\bullet} \, [10^6 \, M_{\odot}] $ & $4.29950$ & 4.1 & 4.8\\[2pt] 
    $ R_0 \, \rm [10^3 \, pc]$ & 8.27795 & 8.1 & 8.9\\ [2pt] 
    $\Lambda$ & 0.001 & 0 & 1 \\
    \hline
\end{tabular}
\end{table}
\begin{table}
\caption{Gaussian priors used in the MCMC analysis. Initial guesses $\Theta_i^0$ coincide with the best-fit parameters found by \textbf{minimize}. $\xi$ and $\sigma$ represent the mean and the standard deviation of the distributions, respectively, and they come from \citet{2015MNRAS.453.3234P}.} 
\begin{tabular}{lccc}
    \hline
    Parameter & $\Theta_i^0$ & $\xi$ & $\sigma$ \\
    \hline
    $ x_0 \, \rm [mas]$ & -0.244 & -0.055 & 0.25 \\ [2pt] 
    $ y_0 \, \rm [mas]$ & -0.618 & -0.570 & 0.15 \\ [2pt] 
    $ v_{x_0} \, \rm [mas/yr]$ & 0.059 & 0.063 & 0.0066 \\ [2pt] 
    $ v_{y_0} \, \rm [mas/yr]$ & 0.074 & 0.032 & 0.019 \\[2pt] 
    $ v_{z_0} \, \rm [km/s]$ & -2.455 & 0 & 5 \\[2pt] 
    \hline
\end{tabular}
\end{table}

The initial points $\Theta_i^0$ in the MCMC are chosen such that they minimise the $\chi^2$ when $f_{\rm SP} = 1$ and $\Lambda = 0$. The minimisation is performed using the Python package \textbf{lmfit.minimize} \citep{lmfit2014} with Levenberg-Marquardt method. 

In the sampling phase of the MCMC implementation, we used 64 walkers and $10^5$ iterations. Since we started our MCMC at the minimum found by \textbf{minimize} we skipped the burning-in phase and we used the last $80\%$ of the chains to compute the mean and standard deviation of the posterior distributions. The convergence of the MCMC analysis is assured by means of the auto-correlation time $\tau_c$, i.e. we ran $N$ iterations such that $N \gg 50 \, \tau_c$.

\section{Corner plots}
\label{app:corner_plot}
Here we report the corner plots for two representative values of $\alpha$ ($\alpha = 0.01$ and $\alpha = 0.001$), to show the behaviour of the parameters when the cloud is located in and outside S$2$'s orbital range. The strong correlation between $\Lambda$ and the periastron passage $t_p$ when $\alpha = 0.01$ can be understood following the argument of \cite{Heissel:2021pcw}: the presence of an extended mass will induce a retrograde precession in the orbit that will result in a positive shift of the periastron passage time, needed to compensate the (negative) shift in the initial true anomaly. Indeed, when considering the Schwarzschild precession, which instead induces a prograde precession (hence a positive initial shift in the true anomaly), $t_p$ will undergo a negative shift, as can be seen from the strong anti-correlation between $f_{\rm SP}$ and $t_p$ reported in \cite{GRAVITY:2020gka}.

\begin{figure*}
\includegraphics[width=\textwidth]{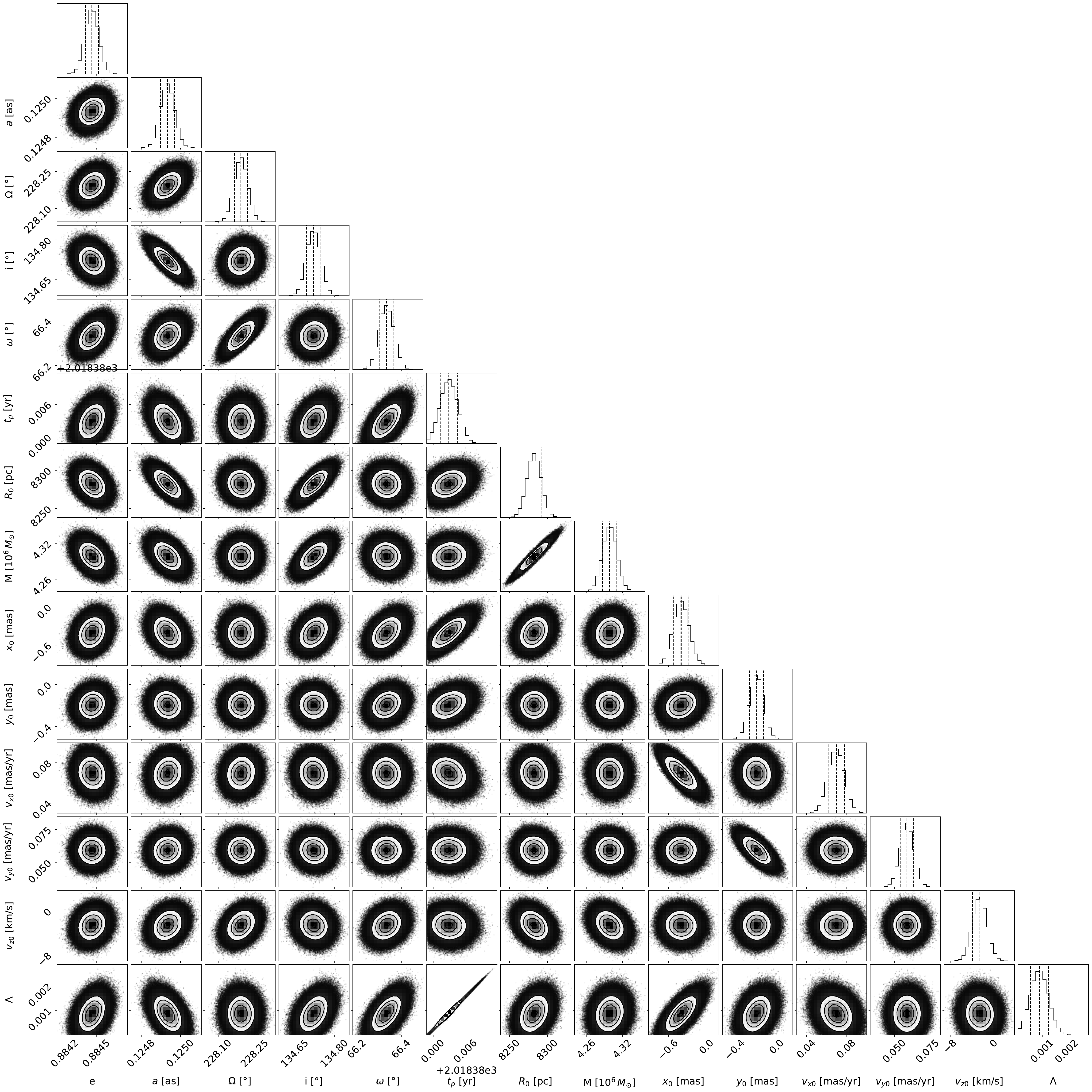}
    \caption{Corner plot of the fitted parameters with $f_{\rm SP} = 1$ and $\alpha = 0.01$. Dashed lines represent the $0.16$, $0.50$ and $0.84$ quantiles of the distributions.}
    \label{fig:corner_plot_alpha1}
\end{figure*}
\newpage
\begin{figure*}
\includegraphics[width=\textwidth]{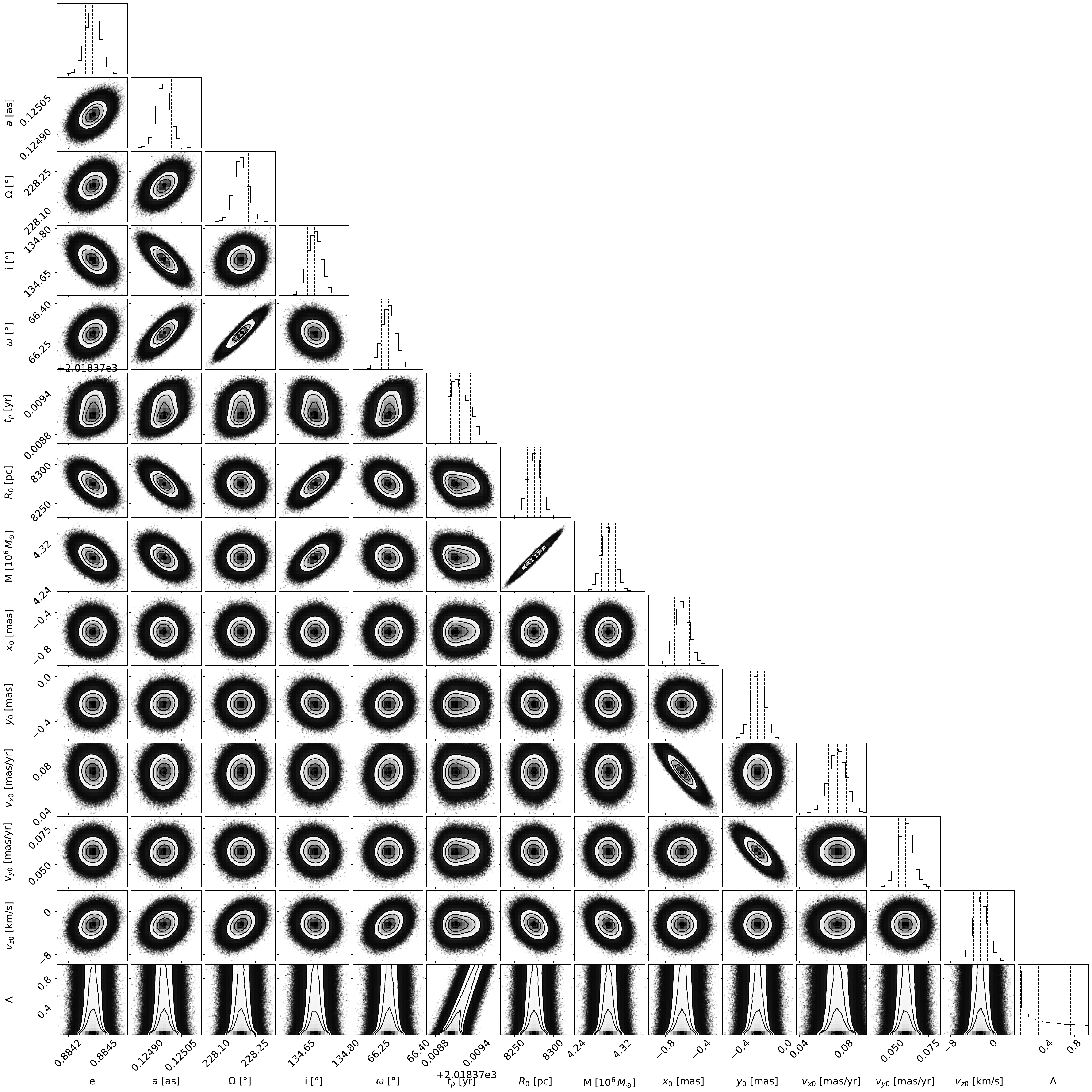}
    \caption{Corner plot of the fitted parameters with $f_{\rm SP} = 1$ and $\alpha = 0.001$. Dashed lines represent the $0.16$, $0.50$ and $0.84$ quantiles of the distributions.}
    \label{fig:corner_plot_alpha2}
\end{figure*}

\bsp	
\label{lastpage}
\end{document}